\documentclass[iop,apj,numberedappendix,twocolappendix,revtex4]{emulateapj}  
\usepackage{amsmath}
\usepackage{bm}
\usepackage{graphicx}
\usepackage{ae,aecompl}
\usepackage[pdfa]{hyperref}
\usepackage{multirow}
\usepackage{color}
\definecolor{blue}{rgb}{0.2,0.2,0.8}
\definecolor{blue}{rgb}{0.2,0.2,0.8}
\definecolor{blue}{rgb}{0.2,0.2,0.8}

\def\eg{{\it e.g.,}}

\def\pmb#1{\setbox0=\hbox{$#1$}%
  \kern-0.25em\copy0\kern-\wd0
  \kern.05em\copy0\kern-\wd0
  \kern-0.025em\raise.0433em\box0}

\long\def\Ignore#1{\relax}

\slugcomment{}

\begin{document}
\title{Stellar Bar evolution in the absence of dark matter halo}
%\shorttitle{Stellar Bar evolution in ...}
\author{Mahmood Roshan }
\affil{Department of Physics, Ferdowsi University of Mashhad, P.O. Box 1436, Mashhad, Iran; \textcolor{blue}{mroshan@um.ac.ir}}
\begin{abstract}
We study the stellar bar growth in high resolution numerical galaxy models with and without dark matter halos. In all models the galactic disk is exponential and the halos are rigid or live Plummer spheres. More specifically, when there is no dark matter halo, we modify the gravitational force between point particles. To do so we use the weak field limit of an alternative theory of dark matter known as MOG in the literature. The galaxy model in MOG has the same initial conditions as in galaxy models with dark matter halo. On the other hand, the initial random velocities and the Toomre's local stability parameter are the same for all the models. We show that the evolution and growth of the bar in MOG is substantially different from the standard cases including dark matter halo. More importantly, we find that the bar growth rate and its final magnitude is smaller in MOG. On the other hand, the maximum value of the bar in MOG is smaller than the Newtonian models. It is shown that although the live dark matter halo may support the bar instability, MOG has stabilizing effects. Furthermore, we show that MOG supports fast pattern speeds, and unlike in the dark matter halo models pattern speed does not decrease with time. Theses differences, combined with the relevant observations, may help to distinguish between dark matter an modified gravity in galactic scales.  
\end{abstract}

\section{\small{Introduction}}
\label{introduction}
It is well-known that bulge-less numerical galactic models which initially are in an equilibrium rotational state, are violently unstable to a pressure dominated system, for example for primary simulations see \citet{miller}; \citet{ho}. This fact is known as the \textit{bar instability} in the literature. More specifically, the disk changes its shape to a rotating bar in a small time scale compared to the rotational period of the outermost particles. However, it was understood that existence of a rigid spherical component can suppress the bar instability \citep{op}. After Ostriker \& Peebles influential work, the role of dark matter halo on the global stability of the galaxies has been extensively investigated, for example see \citet{se81}; \citet{ef};
\citet{at1986}; \citet{at2002} and \citet{se14} for a recent review of the subject.

Although, historically, the dark matter halo introduced to prevent the bar formation, observations show that one-third of disk galaxies are strongly barred. On the other hand, one-third of them are weakly barred, and the remainder do not contain bar. In fact one of the main unsolved problem with these galaxies is the frequency of bars. Several attempts have been done to explain the presence or absence of a bar, for very short review on the subject see \cite{bse16}. In this paper we do not attempt to present a new explanation for the observed bar frequency. We are interested to bar evolution in models when there is no dark matter halo and, instead, the gravitational force is different from the Newtonian case. 

It is important to mention that the bar instability is directly linked to the dark matter problem. On the other hand modified gravity is another path to address this problem. Therefore, the global stability of disk galaxies is an important issue which might help to distinguish between dark matter and modified gravity. In fact, the dark matter particles have not yet been observed while there are several laboratories throughout the world searching for these particles using completely different techniques, see \citet{berton}. This means that modified gravity is still an important approach to address the problem. Theses theories can be divided to two main categories: dark energy models and alternative theories which dispense with the dark matter paradigm. For a review on dark energy models we refer the reader to \cite{cap}. On the other hand Modified Newtonian Dynamics (MOND) is one of the well studied modified theories for dark matter (\citealt{milgrom}; \citealt{fa}). Scalar-tensor-vector theory of gravity known as MOG in the literature is another theory which presented to address the dark matter problem \citep{m2006}. 

Before explaining our aim in this paper, let us briefly mention some well established results on the stability of disk galaxies in both viewpoints, i.e. in particle dark matter and modified gravity. It is well understood that disks in live dark matter halos form bars more readily than in rigid halos \citep{at2002,se16}. In fact angular momentum transfer between the bar and the halo, in principle, can increase the growth rate. This means that the so-called disk stability criteria that derived using rigid halos \citep{op,ef} can  not be used for real galaxies \citep{at2008}. The rotation of the halo can also influence the growth rate. For example, \cite{saha} reported that if the halo rotates in the same direction as the disk, the bar growth will increase. On the other hand, the counter-rotating halos may suppress the instability. 

Another important result of stellar N-body simulations is that the bar pattern speed decreases with time \citep{deb2000,cosmo}. However observations do not confirm this point \citep{pattern}. Therefore it is still a challenge for standard dark matter halo models.

On the other hand, less investigation has been done on the stability of disk galaxies in modified gravity theories. The local stability of disk galaxies has been investigated in \cite{mil89}. The global stability of disk galaxies in MOND has been investigated using N-body simulations in \citet{chris}; \citet{brada} and \citet{ti}. Using a low resolution simulation, it has been shown in \citet{chris} that disk galaxies are more stable in MOND than equivalent models in Newtonian gravity. A similar result has been reported in \cite{brada}. However, advanced simulations with larger number of particles show that the bar growth rate in MOND is larger than in Newtonian models \citep{ti}. This means that galactic disks without dark matter halo in MOND develop a bar sooner than in Newtonian models with dark matter halo. However, in MOND the bar starts to weaken while in the dark matter model it does not stop growing. In this case, the final magnitude of the bar at the end of the simulation in MOND is smaller than the dark matter halo models. 

\cite{ro2015} investigated the local stability of the galactic disks in MOG. In other words, the generalized version of the Toomre criterion has been found and applied to a sample of spiral galaxies. The result shows that there is no significant difference between MOG and the standard picture in the local stability issue. More specifically, the effects of MOG appear at large distances and, naturally, does not affect the dynamics of local perturbations. \cite{rokho} investigated the global stability of the Maclaurin disk in MOG using semi analytic method. On the other hand the global stability of exponential and Mestel disk galaxies in MOG has been investigated in \cite{gr2017} using a low resolution N-body simulations. Although there is no meaningful difference in the local stability of disk galaxies in MOG and the dark matter model, the global stability is substantially different. More specifically, it is found in \cite{gr2017} that disks are more stable in MOG and the growth rate is smaller. 

However the maximum number of particles in the simulations of \cite{gr2017} is $N=2\times 10^4$. Therefore the results may suffer from shot noise and numerical artifacts. Although some tests have been done to ensure the reliability of the results, it still seems necessary to perform high resolution simulations. Furthermore galactic models in \cite{gr2017} do not include dark matter halos. On the other hand, dark matter halo is one of the main components of a spiral galaxy in the standard picture. Therefore, without adding dark matter halo it is not possible to compare appropriately galactic dynamics in MOG with the standard case. Therefore we generalize simulations of \cite{gr2017} by using another N-body code which allows large number of particles. In fact in this paper we will use $2\times 10^6-10^7$ particles for each live component. On the other hand we add the dark matter component and make more realistic galactic models.

The outline of this paper is as follows: In section \ref{wfl} we briefly introduce MOG and discuss its weak filed limit. In section \ref{gi} we describe the numerical method for setting the initial conditions and evolving the point particles. In fact we make three models which have identical initial conditions, one model in MOG without dark matter halo and two models in the standard case including dark matter halo. In section \ref{res} we compare the bar evolution in these three models and discuss the significant differences and their relevance to current observations. Finally we discuss the future developments.

\section{\small{Weak field limit of MOG}}
\label{wfl}

MOG is a relativistic generalization of General Relativity (GR) which beside the metric tensor has two scalar fields, i.e. $G$ and $\mu$, and one massive Proca vector field $\phi^{\alpha}$. In other words MOG is a scalar-tensor-vector theory of gravity. In particle physics language, MOG possesses a spin two graviton, i.e. the metric tensor $g_{\mu\nu}$, a spin one massive graviton, namely the vector field $\phi^{\alpha}$, and a spin zero massless graviton, i.e. the scalar field $G$.

Existence of these fields enables the theory to address the dark matter problem without invoking dark matter particles. More specifically, in the weak field limit, MOG explains the flat rotation curve of spiral galaxies \citep{br2006}, as well as the mass discrepancy in galaxy clusters \citep{br2006b}. This theory has been applied also to Bullet Cluster and Train Wreck Cluster in order to explain the unusual lensing data, see \citet{im}. On the other hand the cosmological aspects of the theory has been widely investigated, for example see \citet{m2009}; \citet{m2015}; \citet{roepjc}; \citet{jamali}; \citet{shojai} and \cite{jamali2}. It is important mentioning that MOG has a viable sequence of cosmological epochs. On the other hand, in its original form, it can not be considered as a dark energy model. In other words, this theory is designed to address the dark matter problem and uses a cosmological constant $\Lambda$ to be consistent with the dark energy observations.

In order to study the dynamics of a disk galaxy, the weak field limit of the theory is required. The weak field limit of MOG has been studied in
\citet{m2013} and \citet{ro2014}. {In the following we briefly review the weak field limit. The field equations of the theory, which are obtained from varying the generic action of the theory with respect to the fields, i.e. the metric tensor $g_{\mu\nu}$, $G$, $\phi_{\alpha}$, $\mu$, are given by} 
 \begin{equation}
 \begin{split}
&G_{\mu\nu}+g_{\mu\nu}\Big(\frac{2}{G^2}\nabla_{\alpha}G\nabla^{\alpha}G-\frac{\square G}{G}+G\omega\mu^2 \phi_{\alpha}\phi^{\alpha}\\&~~~-4\pi\frac{\nabla_{\alpha}\mu\nabla^{\alpha}\mu}{\mu^2} \Big)-\frac{2\nabla_{\mu}G\nabla_{\nu}G}{G^2} + \frac{\nabla_{\mu}\nabla_{\nu}G}{G}\\&~~~-8\pi\Big(\frac{1}{4\pi}G\omega\mu^2\phi_{\alpha}\phi^{\alpha}-\frac{\nabla_{\mu}G\nabla_{\nu}G}{G^2}-
\frac{\nabla_{\mu}\mu\nabla_{\nu}\mu}{\mu^2}\Big)\\&~~~
+\frac{G\omega}{4\pi}\Big(B^{\alpha}_{~\mu}B_{\nu\alpha}+\frac{1}{4}B^{\alpha\beta}B_{\alpha\beta}g_{\mu\nu}\Big)=-8\pi G T_{\mu\nu}
\end{split}
\label{mog11}
\end{equation}
 \begin{equation}
    \square G+\frac{G}{2} \Big(\frac{\nabla^{\alpha}\mu\nabla_{\alpha}\mu}{\mu^2}+\frac{R}{8\pi}\Big)
    -\frac{3}{2G}\nabla^{\alpha}G\nabla_{\alpha}G=0
 \label{mog13}
  \end{equation}
 \begin{equation}\label{mog12}
    \omega \nabla_{\mu}B^{\mu\nu}+\omega \mu^2\phi^{\nu}=4\pi J^{\alpha}
 \end{equation}
 \begin{equation}\label{mog13}
    \square \mu-\frac{\nabla_{\alpha}\mu\nabla^{\alpha}\mu}{\mu}-\frac{\nabla_{\alpha }\mu\nabla^{\alpha}G}{G}+\frac{1}{4\pi}G\omega\mu^3\phi^{\alpha}\phi_{\alpha}=0
 \end{equation}
{where $\square=\nabla_{\mu}\nabla^{\mu}$, $B_{\mu\nu}=\nabla_{\mu}\phi_{\nu}-\nabla_{\nu}\phi_{\mu}$, $G_{\mu\nu}$ is the Einstein tensor, $R$ is the Ricci scalar, $T_{\mu\nu}$ is the energy-momentum tensor, $\omega$ and $\kappa$ are coupling constants, $J^{\alpha}$ is a "matter current" obtained from the variation of the matter action with respect to the vector field, and its time component is written as $J^0=\kappa \rho$. Only in this section, we use units in which the speed of light is unity. In order to find the weak field limit, let us perturb the Minkowski metric $\eta_{\mu\nu}$ as follows}
 \begin{equation}
 g_{\mu\nu}=\eta_{\mu\nu}+h_{\mu\nu}=\left(
              \begin{array}{cc}
                1+2 \Phi & 0 \\
                0& (-1+2\Psi)\delta_{ij}\\
              \end{array}
            \right)
  \label{pert1}
 \end{equation}
{where $|h_{\mu\nu}|\ll 1$. On the other hand, using the field equations it is easy to show that the background values of $J^{\alpha}$ and $\phi^{\alpha}$ are zero. Therefore, the linear perturbation of other fields and the matter current can be written as}
  \begin{equation}
 \begin{split}
 & \phi^{\mu}=\phi_1^{\mu}(\mathbf{r},t), ~~~~~G=G_{0}+G_1 (\mathbf{r},t)\\&
            \mu=\mu_{0}+\mu_1(\mathbf{r},t)~~~~~~~            J^{\mu}=J_1^{\mu}(\mathbf{r},t)
 \end{split}
 \label{pert2}
 \end{equation}
 {where subscripts "$0$" and "$1$" stand for background and perturbed quantities respectively. Now, assuming a perfect fluid energy-momentum tensor and substituting equations \eqref{pert1} and \eqref{pert2} into the field equations \eqref{mog11}-\eqref{mog13}, we find the linearised field equations}
\begin{equation}
\nabla^2 \Psi=4\pi G_0 \left(\frac{16\pi-2}{16\pi-3}\right)\rho
\label{jmog6}
\end{equation}
\begin{equation}
\nabla^2 \Phi=4\pi G_0 \left(\frac{16\pi}{16\pi-3}\right)\rho
\label{jmog60}
\end{equation}
\begin{equation}
\nabla^{2}\phi_1^{\mu}-\mu_0 \phi_1^{\mu}=-\frac{4\pi}{\omega} J_1^{\mu}
\label{jmog10}
\end{equation}
\begin{equation}
\nabla^2 G_1=\frac{8\pi G_0^2}{16\pi-3}\rho
\label{jmog7}
\end{equation}
\begin{equation}
\nabla^2 \mu_1=0
\end{equation}
{Now in order to find the modified version of the Poisson equation, we need to linearise the equations of motion of a test particle, or equivalently the geodesic equation, in MOG. the geodesic equation in MOG differs from GR and is given by }
\begin{equation}
\frac{d^2x^i}{d\tau^2}+\Gamma^{i}_{\alpha\beta}\frac{dx^{\alpha}}{d\tau}\frac{dx^{\beta}}{d\tau}=\kappa B^{i}_{~\alpha}\frac{dx^{\alpha}}{d\tau}
\label{jmog9}
\end{equation}
{in which $\Gamma$s are the Christoffel symbols. It is easy to show that in the weak field limit this equation takes the following form } 
\begin{equation}
\frac{d^2\mathbf{r}}{dt^2}\simeq-\nabla \Phi_{\text{eff}}
\label{new0}
\end{equation}
{where the effective potential $\Phi_{\text{eff}}$ is defined as}
\begin{equation}
\Phi_{\text{eff}}=\Phi+\kappa \phi_1^{0}
\label{jmog12}
\end{equation}
{On the other hand, equations \eqref{jmog60} and \eqref{jmog10} can be straightforwardly solved as}
\begin{equation}
\Phi=-\frac{16\pi G_0  }{16\pi-3}\int \frac{\rho(\mathbf{r'})}{|\mathbf{r}-\mathbf{r'}|}d^3x'
\end{equation}
\begin{equation}
\phi_1^{0}=\frac{\kappa}{\omega}\int \frac{e^{-\mu_0|\mathbf{r}-\mathbf{r'}|}}{|\mathbf{r}-\mathbf{r'}|}\rho(\mathbf{r'})d^3x'
\label{jmog11}
\end{equation}
{It is clear from Yukawa equation \eqref{jmog10} that the scalar field $\mu$ appear as a mass for the vector field. Using these solutions, we rewrite equation \eqref{jmog12} as }
\begin{equation}
\Phi_{\text{eff}}(\mathbf{r})=-G_N\int \frac{\rho(\mathbf{r'})}{|\mathbf{r}-\mathbf{r'}|}\Big(1+\alpha-\alpha e^{-\mu_0|\mathbf{r}-\mathbf{r'}|}\Big)d^3x'
\label{effp}
\end{equation}
{where $\alpha$ is a dimensionless parameter defined as $\alpha=\frac{\kappa}{\omega G_N}$,   and $G_N$ is the Newton's gravitational constant. Also, in order to recover the Newtonian gravity in small distances, it is necessary to set $16\pi G_0/(16\pi-3)-\kappa^2/\omega=G_N$. Now using equations \eqref{jmog60}, \eqref{jmog10}, \eqref{jmog12} and \eqref{effp}, one can readily verify that the Poisson equation in Newtonian gravity is replaced with} 
\begin{equation}
 \nabla^2 \Phi_{\text{eff}}(\mathbf{r})=4\pi G_N \rho +\alpha \mu_0^2 G_N \int \frac{e^{-\mu_0 |\mathbf{r}-\mathbf{r'}|}}{|\mathbf{r}-\mathbf{r'}|} \rho(\mathbf{r'})d^3x'
 \label{n1}
\end{equation}

{Hereafter, for simplicity, we drop the subscript "$N$" in the Newton's constant $G_N$. As it is clear we have two free parameters $\alpha$ and $\mu$ which should be fixed using relevant dark matter observations. From rotation curve data of spiral galaxies we have $\alpha=8.89 \pm 0.34$ and $\mu_0=0.042 \pm 0.004 ~\text{kpc}^{-1}$ \citep{m2013}. } 

{Now let us find the gravitational potential of a test particle $M$ located at $\mathbf{r'}=0$. In this case the density is given by $\rho(\mathbf{r'})=M \delta(\mathbf{r'})$, where $\delta$ is the Dirac delta function. By substituting this density into equation \eqref{effp} we arrive at}
\begin{equation}
 \Phi_{\text{eff}}(r)=-\frac{G M}{r}[1+\alpha-\alpha e^{-\mu_0 r}].
\label{potmog}
 \end{equation}
 {Consequently, the gravitational force experienced by another unit mass located at distance $r$ from the point mass $M$ is given by}
\begin{equation}
\mathbf{F}=-\frac{G M}{r^2}[1+\alpha-\alpha(1+\mu_0 r)e^{-\mu_0 r}]\frac{\mathbf{r}}{r}
 \label{accmog}
\end{equation}
{At small radii the Newtonian force is recovered. On the other hand at larger scales the Yukawa term is suppressed and the the gravitational force in MOG is enhanced by factor $(1+\alpha)$ compared to the Newtonian force. }

As in other alternative theories of  gravity for dark matter, the gravitational force is stronger than the Newtonian force. It is interesting to mention that in strong field limit, MOG predicts new effects which are absent in GR. For example, MOG leads to \textit{gravito-electrical} and \textit{gravito-magnetic} effects which are directly related to the existence of the vector field, for more details see \cite{romero} and \cite{gravitoelectric}. However, we do not expect this effect in the galactic scale where almost all the dynamics can be explained in the Newtonian limit.
\begin{deluxetable*}{ccccccccccccccc}
\tablecolumns{15}
\tablecaption{Model information}
\tablehead{
& \colhead{Disk} & \colhead{Disk} & \colhead{Disk} & \colhead{Initial}  & \colhead{Halo} & \colhead{Halo} & \colhead{Halo}&\colhead{Halo}&\colhead{Time}&\colhead{Cylindrical}&&\colhead{Spherical}&\colhead{Softening} \\
\colhead{Run} & \colhead{Type} & \colhead{Edge} & \colhead{Mass}&\colhead{$Q$}&\colhead{Type}&\colhead{Mass}& \colhead{Scale} & \colhead{$r_{\rm max}$}&\colhead{Step}&\colhead{Polar grid}&$\delta z$&\colhead{Grid}&\colhead{Length}&
}
\startdata\\
EN\Ignore{200} & Exp&4 & 1 & 1.5 & none & ... & ... & ... & 0.01 & $390\times 407\times 135$&0.08&...&0.16&\\
EM\Ignore{200} & Exp&4 & 1 & 1.5 & none & ... & ... & ... & 0.01 & $193\times 224\times 125$&0.08&...&0.16&\\
EPR\Ignore{200} & Exp&4 & 1 & 1.5 & Plum & 7.5 & 5.5 & ... & 0.01 & $99\times 128\times 125$&0.08&...&0.16&\\
EPL\Ignore{200} & Exp&4 & 1 & 1.5 & Plum & 8 & 10 & 30 & 0.01 & $100\times 128\times 125$&0.08&1001&0.16&\\
\tablecomments{Column 1: Letter identification for the simulation. Column 2: Type
   of initial disk; ``Exp'' is exponential disk. Column 3:
   Initial outer radius of the disk in units of $R_d$.  Column 4:
   Initial disk mass $M_d$.  Column 5: Initial Toomre's $Q$ parameter.  Column 6: Type of halo, ``Plum'' is the Plummer isotropic sphere.  Column 7: Mass of the halo
   component $M_b$.  Column 8: Radial scale of the halo
   component $b$ in units of $R_d$. Column 9: Outer edge of halo, $r_{\rm
     max}$ in units of $R_d$. Column 10: the basic time step in units of $\tau_0$. Column 11: number of rings, spokes, and planes in the cylindrical polar grid. Column 12: vertical distance between grid planes. Column 13: the number of shells in the spherical grid. This grid extends to $R=180 R_d$. Column 14: the gravity softening length in units of $R_d$.}.
 \label{tab1}
\end{deluxetable*}

\section{\small{Numerical method}}
\label{gi}
We use the N-body code developed by \cite{se2014}. For galaxy model with live halo, we use the hybrid N-body code described by \citep[Appendix B]{se3} in which two grids are used to compute the gravitational field: a 3D cylindrical polar grid for the disk component, and a spherical grid for the live halo component. One may find the full description of the code in the online manual \citep{se2014}. 

For galaxy models in MOG, we have modified the code in order to take into account the modified gravity corrections to the gravitational force. In our galaxy model in MOG, as expected, there is no dark matter halo. The energy conservation has been tested, and the error in total energy is always smaller than $1$\%. 
\subsection{Initial conditions}

All the models we present in this paper start with an exponential disk given by the following surface density
\begin{eqnarray}
\Sigma(R)=\frac{M_{d}}{2\pi R_d^2}\,e^{-R/R_d}
\end{eqnarray}
where $M_d$ is the disk mass, and $R_d$ is its length scale. The extend of the disk is limited by tapering the surface density with a function that varies as a cubic polynomial from unity at $R=3.2\, R_d$ to zero at $R=4 \,R_d$. Using the epicyclic approximation, in which particles move on nearly circular orbits, we set the radial velocity dispersion $\sigma_R$ such that to ensure about the local stability of the disk. In fact the local stability criterion is written as $Q>1$ where the Toomre's stability parameter $Q$ \citep{toom64} is given by 
\begin{eqnarray}
Q(R)=\frac{\sigma_R}{\sigma_{R,\text{crit}}}~~~~~~\text{where} ~~~~\sigma_{R,\text{crit}}=\frac{3.36 G \Sigma}{\kappa}
\end{eqnarray}
and $\kappa(R)$ is the epicyclic frequency. Using the epicycle relation, we also find the initial azimuthal velocity dispersion as $\sigma_{\phi}=\kappa \sigma_R/2\Omega$, where $\Omega(R)=v_c/R$ and $v_c$ is the circular velocity. On the other hand the disk has a Gaussian density profile in the vertical direction with the scale height $z_0=0.05 R_d$. Furthermore, the vertical velocities are determined by integrating the vertical one dimensional Jeans equation as
\begin{eqnarray}
\sigma_z^2(R,z)=\frac{1}{\rho(R,z)}\int_z^{\infty}\rho(R,z')\frac{\partial \Phi}{\partial z'}dz'
\end{eqnarray}

For the halo component, we have chosen the isotropic Plummer sphere, which has the following density profile
\begin{eqnarray}
\rho(r)=\frac{3 M_h}{4\pi b^3}\Big[1+\Big(\frac{r}{b}\Big)^2\Big]^{-5/2}
\end{eqnarray}
where $M_h$ is the halo mass and $b$ is the characteristic radius. The values of these parameters are selected so that the Newtonian model has the same rotation curve as the disk in modified gravity (which has no spherical component). The halo is truncated at $r=3\,b$. It is important mentioning that this mass distribution is consistent with the rising rotation curves in late type galaxies of lower luminosity \citep[\eg][]{b2008}. 

{It is necessary to mention that we have two main reasons for choosing this halo. In fcat with this choice, it is easy to match the rotation curves of the dark matter and modified gravity models. On the other hand } this profile has been used also in \cite{com1} in order to compare bulge formation in MOND and standard Newtonian dynamics. Also it has been used in \cite{ti} to study the stellar bar evolution in MOND. {Therefore this choice of halo enables us to make a direct comparison with  \cite{ti}, and conseuently a direct comarison between MOND and MOG. }

 \cite{de} gives expressions for the distribution function (DF) of the Plummer sphere, characterized by a parameter $q$. We set $q=0$ which correspond to a isotropic sphere. Therefore we start with a sphere with known DF when it is in isolation. However, when a disk component is added, one needs to find the equilibrium DF of the spherical component in the composite halo and disk model. To do so we use a halo compression algorithm described by \cite{y1980} and \cite{sm2005}. This algorithm uses the fact that both radial action and the total angular momentum are conserved during compression. Finally the particles will be selected from the new DF function as described in \cite{deb2000}. For more details see \cite{sm2005} in which the iterative solution for a single spheroid with an added disk were studied using the above mentioned procedure. 
 
We emphasize that no bulge is included in our models. More specifically, we study two single component models "EN" and "EM" with an exponential disk. EN is the exponential disk model in Newtonian gravity and EM is the corresponding model in MOG. On the other hand "EPR" and "EPL" are two component models including an exponential disk and a Plummer halo. In EPR the halo is rigid, and in EPL is live and responsive. The properties of these models have been summarized in Table \ref{tab1}. For each responsive components we employ two million particles. For example, in EPL we employed two million particle for the disk and two million particle for the live halo. Other models have two million particles. In order to check that the results do not depend on the number of particles, {we also increase the number of particles by a factor of $10$ in all the models}.

\subsection{Units}
We use $M_d$ as the unit of mass and $R_d$ as the unit of length throughout of this paper. In other words we use units such that $R_d=M_d=1$. Also we scale the Newtonian gravitational constant as $G=1$. In this case, the unit of velocity is $V_0=(G M_d/R_d)^{1/2}$, and time unit is $\tau_0=R_d/V_0=(R_d^3/G M)^{1/2}$. For example a suitable choice for our models can be $R_d=2.6$ kpc and $\tau_0=10$ Myr, which yields $M_d\simeq 4 \times 10^{10} M_{\odot}$ and $V_0\simeq 254\, \text{km}/\text{s}$. For all the models we compute the evolution for $500 \tau_0$. We use a time step $\Delta t= 0.01 \tau_0$ and the time step is increased by successive factors of 2 in three radial zones. The cubic spline softening length \citep{mon1992} $\epsilon$ is set to $0.16\,R_d$ in all the models. Also, except in the model EPR which has a rigid component, the grid is recentered every $16$ time step. This dynamic recentering helps to avoid numeric artifacts.
\subsection{Rotation curves}
\begin{figure} 
 \centerline{\includegraphics[width=8.4cm]{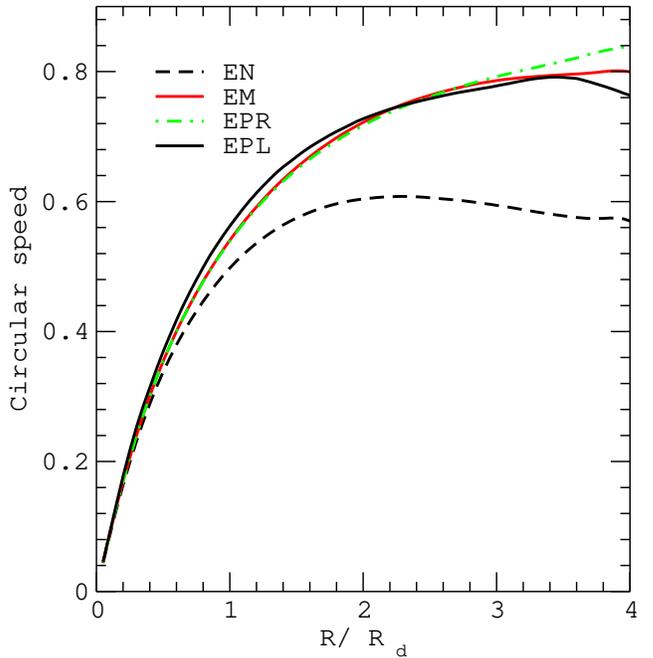}}
\caption[]{{Initial rotational velocities for our four models. The dashed curve belongs to model EN which is a single exponential disk in Newtonian gravity. The red curve is the corresponding single component model in MOG with free parameters $\alpha=8.9$ and $\mu_0=0.015 R_d^{-1}$. The dot dashed and black solid curves belong to tow component models EPR and EPl respectively.}}
\label{vc}
\end{figure}

Using the above mentioned units, the rotation curves of our four models have been illustrated in Fig. \ref{vc}. It is necessary to mention that for an exponential disk the rotation curve in MOG is computed for a given $\alpha$ and $\mu_0$ first, i.e. model EM, and then equivalent Newtonian systems are built with a dark matter halo added, in order to obtain the same rotation curve. More specifically by changing the halo parameters, i.e. mass and the characteristic length, in both the rigid and live cases, we construct models EPR and EPL which their rotation curve is similar to the model EM. It should be emphasized that in order to compare the stellar bar evolution in MOG and Newtonian gravity, it is necessary to start with the same initial rotation curves. In this case, the dispersion velocities are also almost the same. One should note that in model EM we have set $\alpha=8.9$ and $\mu_0=0.015 R_{d}^{-1}$, which $\alpha$ is close to the observed value. However, we are constructing toy galactic models and naturally we do not have to use the observational values. On the other hand, we know that these parameters change from galaxy to galaxy. 
\subsection{Bar amplitude}

As we mentioned before, the main aim of this paper is to study and compare the bar growth in Newtonian gravity and MOG. Therefore, the main quantity to be measured in our simulations is the bar amplitude. It is important to mention that the code determines the gravitational forces on the cylindrical polar grid using sectoral harmonics $0\leq m\leq 8$, and surface harmonics $0\leq l\leq 4$ on the spherical grid. For more details see the online manual available at \url{http://www.physics.rutgers.edu/~sellwood/manual.pdf}.  

The amplitude of the non-axisymmetric features can be measured by computing
\begin{equation}
A_m(t)=\sum_j \mu_j e^{im\phi_j(t)}
\label{ex}
\end{equation}
where $\mu_j$ is the mass of the  $j$th particle particle and $\phi_j$ is its cylindrical polar angle at time $t$. One should note that this summation is over the disk particles only. So the bar amplitude, for which $m=2$, would be obtained by the ratio $A_2/A_0$.

\begin{figure}
 \center
\centerline{\includegraphics[width=8.5cm]{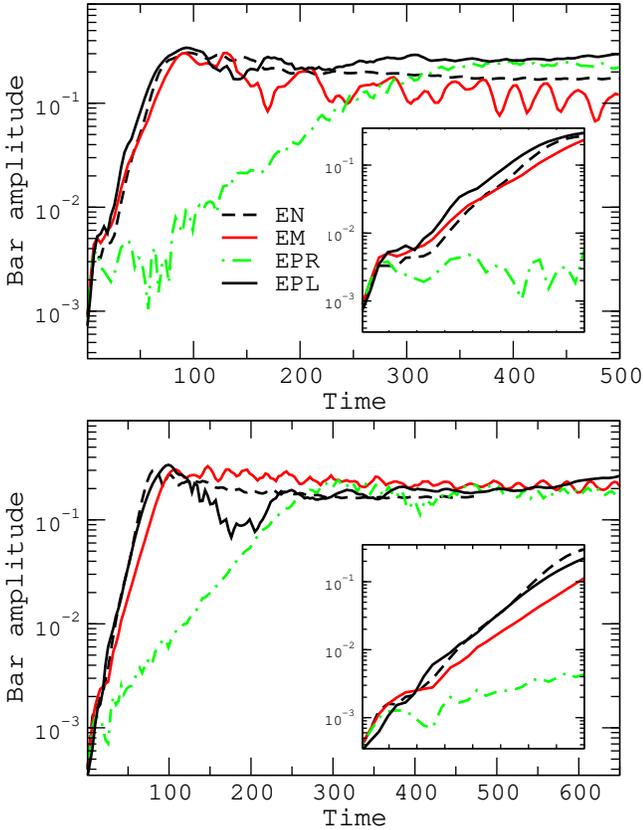}}
\caption[]{{The top panel shows the evolution of the bar amplitude for our models when each responsive component has $2\times 10^6$ particles. The bottom panel shows the bar amplitude when {there are $2\times 10^7$ particles in each responsive component. For example the total number of the particles in model EPL is $4\times 10^7$. The small boxes in each panel show the evolution of the bar at early times, i.e. $\tau<100$.}}}
\label{bar}
\end{figure}

\section{\small{Results}}
\label{res}
\subsection{Stabilizing effects of MOG}
In the recent paper \cite{gr2017}, we have compared two galaxy models which are similar to EM and EN. In fact without adding a spherical halo, it has been shown that MOG has stabilizing effects. More specifically, increasing the MOG free parameters, supports the global stability of the exponential disk. Albeit this point has been checked also for Mestel disk. In fact, the procedure in \cite{gr2017} is similar to what presented in \cite{op}, in which it has been proved that a rigid halo has stabilizing effects. The maximum number of particles in \cite{gr2017} is $2\times 10^4$. Therefore the results may suffer from numeric artifacts caused by employing small number of particles. Here we have increased the particle number to $2\times 10^7$ and used a code which employs a completely different technique to compute the gravitational field. In model EN, the disk rapidly extends to larger radii. Therefore we have to enlarge the grid in order to avoid particles to escape from the grid. In the following let us briefly show that the main results of \cite{gr2017} do not change by increasing the number of particles. Then we will discuss more realistic models by adding dark matter halos to the Newtonian models.

In the model EM we have chosen $\alpha=8.9$ and $\mu_0=0.015 R_d^{-1}$. The corresponding rotation curve is shown in Fig. \ref{vc}. As it is expected the circular velocity at large distances from the center of the disk is higher than the Newtonian model EN. The evolution of bar amplitude has been shown in Fig. \ref{bar}. In both top and bottom panels in this figure, the black dashed curve belongs to model EN and the red solid curve belongs to model EM. {We have used a logarithmic scale so that the growth rate can be estimated from the slop of the curve in the period of linear growth. The small boxes inside each panel show the evolution of the bar amplitude at early times $\tau< 100$.} It is clear from the top panel that the exponential instability growth starts around $\tau\simeq 25$ in both models EN and EM. However, in the model EM the growth rate is smaller than the Newtonian case. This point is more clear from the bottom panel where we have employed larger number of particles. In both cases, after reaching a maximum the bar decays slowly at large times. The slop of this decrease is smaller in the bottom panel. However the qualitative behavior is almost the same. It is clear from the top panel that the final magnitude of the bar in EM is smaller than EN. In the bottom panel this will happen at larger times. This fact has also been reported in \cite{gr2017}. 

\begin{figure} 
\centerline{\includegraphics[width=8.5cm]{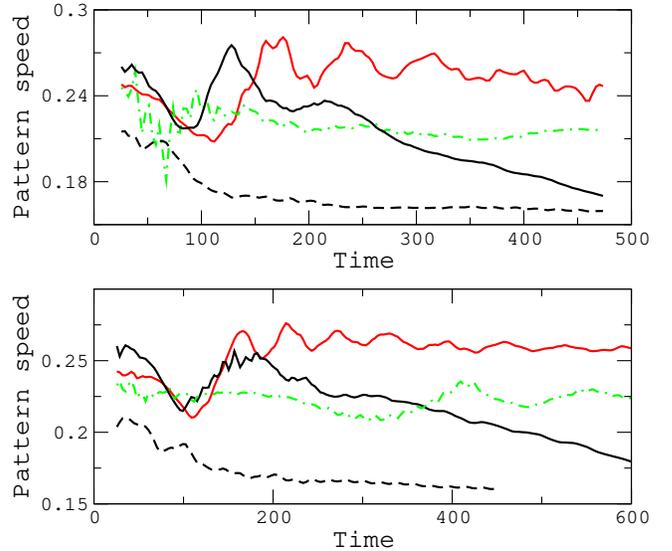}}
\caption[]{{{The pattern speed $\Omega_p(t)$ with respect to time for the models. In the top panel there are $10^6$ particles in each live component , and in the bottom panel there are $2\times 10^7$ particles.}}}
\label{ps}
\end{figure}

The bar pattern speed has been shown in Fig. \ref{ps}. Similar to the results presented in \cite{gr2017}, the pattern speed in EM is larger than EN model. 

 \begin{figure*}
\centering\includegraphics[width=8cm]{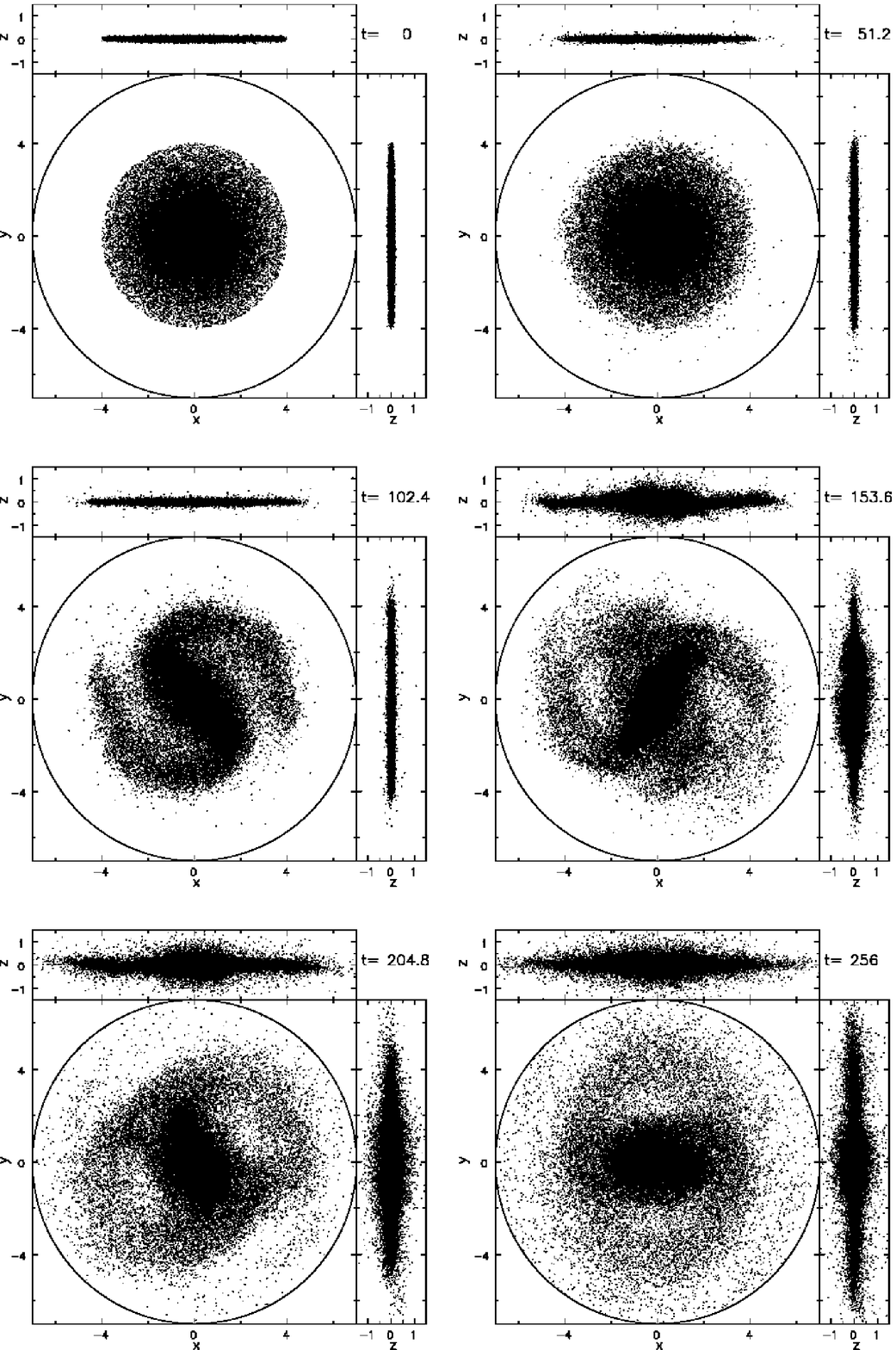}\hspace{0.5 cm} \includegraphics[width=8cm]{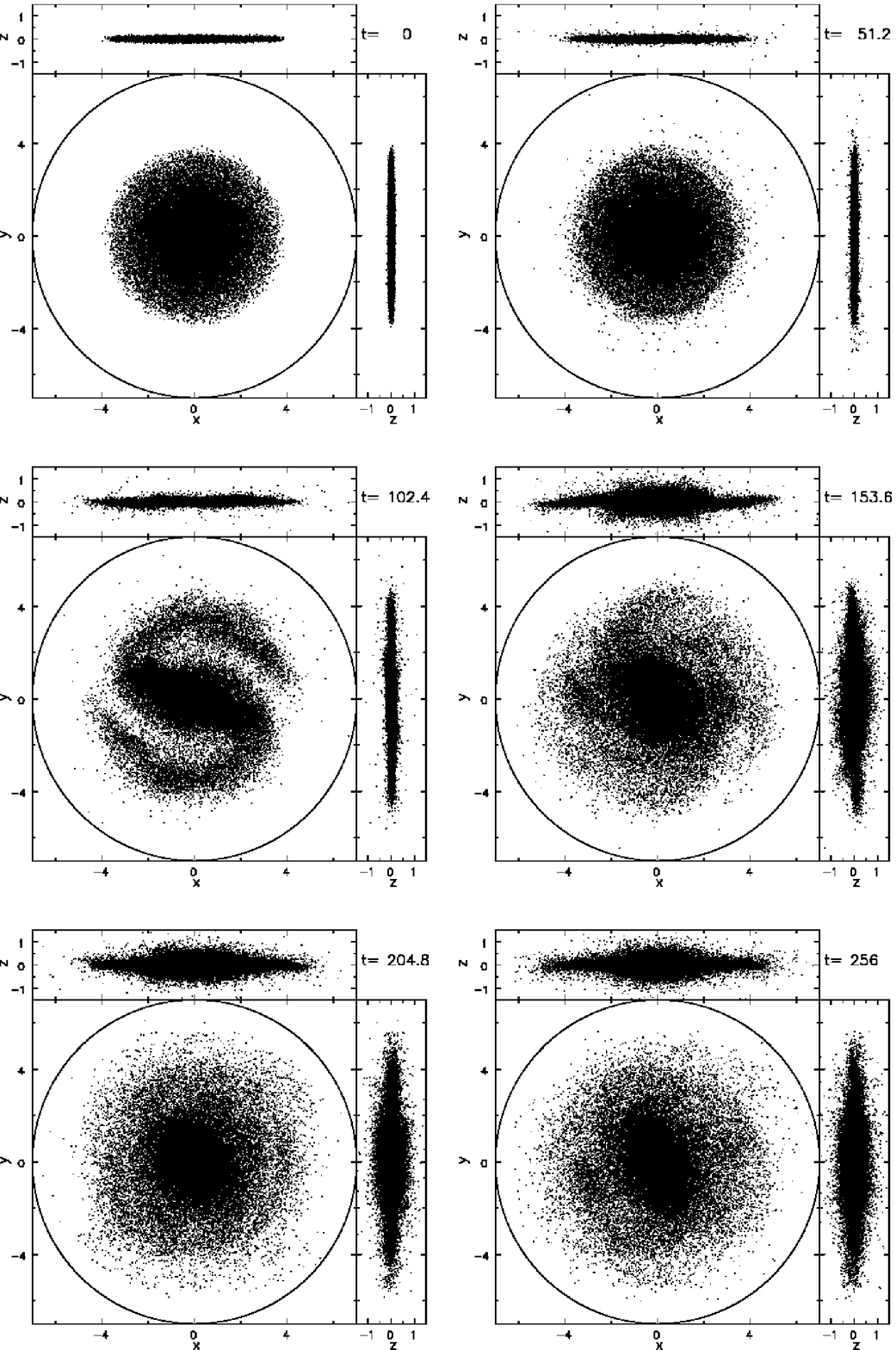}\vspace{0.5 cm}
\caption[]{{{The evolution of the disk, with $2\times 10^7$ particles in each live component, in models EM (left panel) and EPL (right panel) projected on the $x-y$, $x-z$ and $y-z$ planes. Only $5\times 10^4$ particles have been shown in each panel.}}}
\label{pos}
\end{figure*}

\subsection{Bar growth}

As mentioned before it is necessary to compare MOG with Newtonian models which include dark matter halo. In this section we compare model EM with two component models EPR and EPL. EPR has a rigid Plummer halo and EPL includes a live Plummer halo. In fact we have chosen the halos physical properties such that the rotation curve coincides with that of the model EM. As it is clear from the Table \ref{tab1}, the mass and the characteristic length scale in the rigid halo is different from the live halo. In the case of live halo, if we use the same parameters used in the rigid halo, the rotation curves will not match. In fact in model EPL the halo is compressed by the growth of the initial disk and leads to a different rotation curve than that of model EPR. For example see models "GR" and "GL" in \cite{bse16}.
 \begin{figure*}
\centerline{\includegraphics[width=8cm]{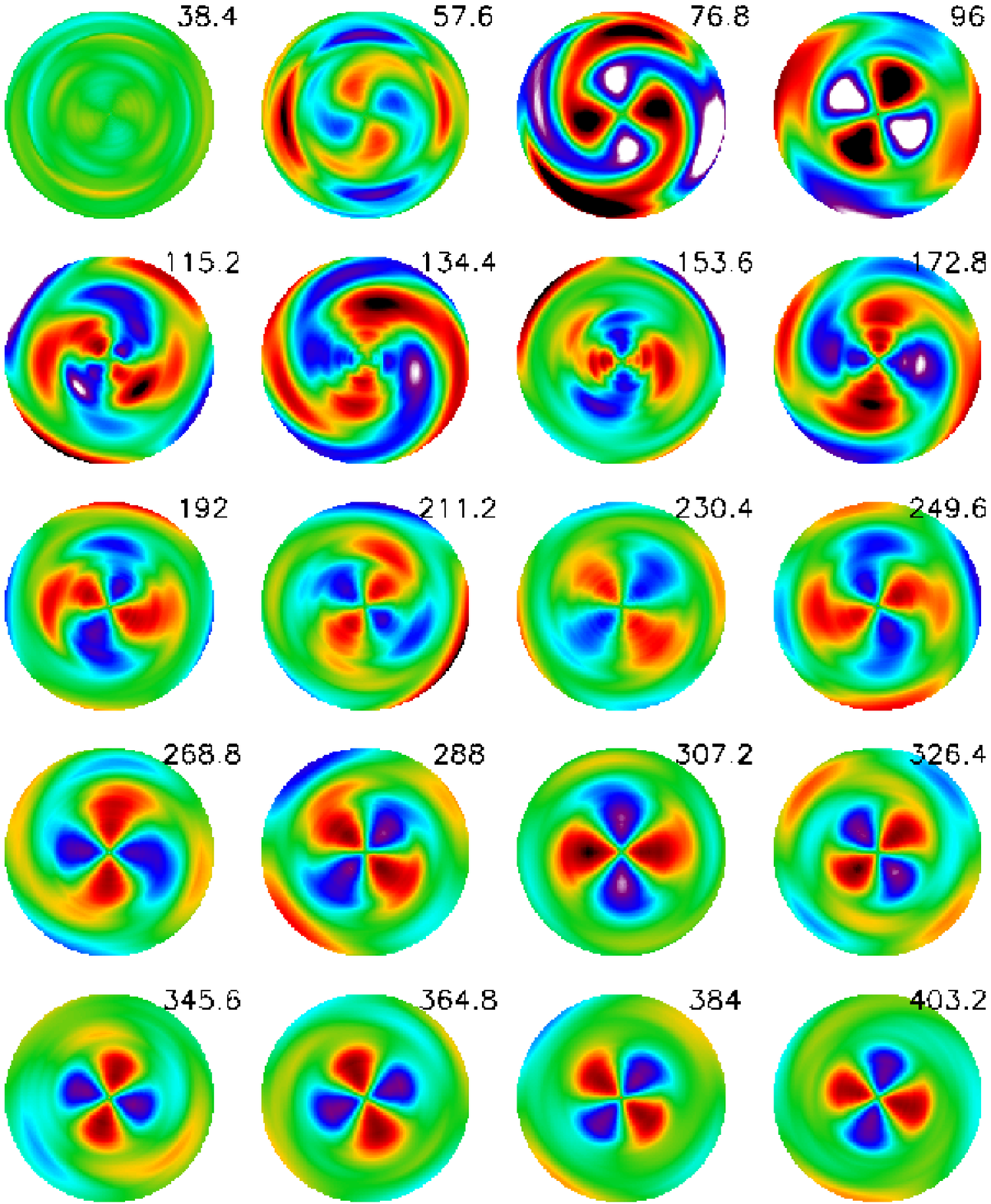}\hspace{1 cm} \includegraphics[width=8cm]{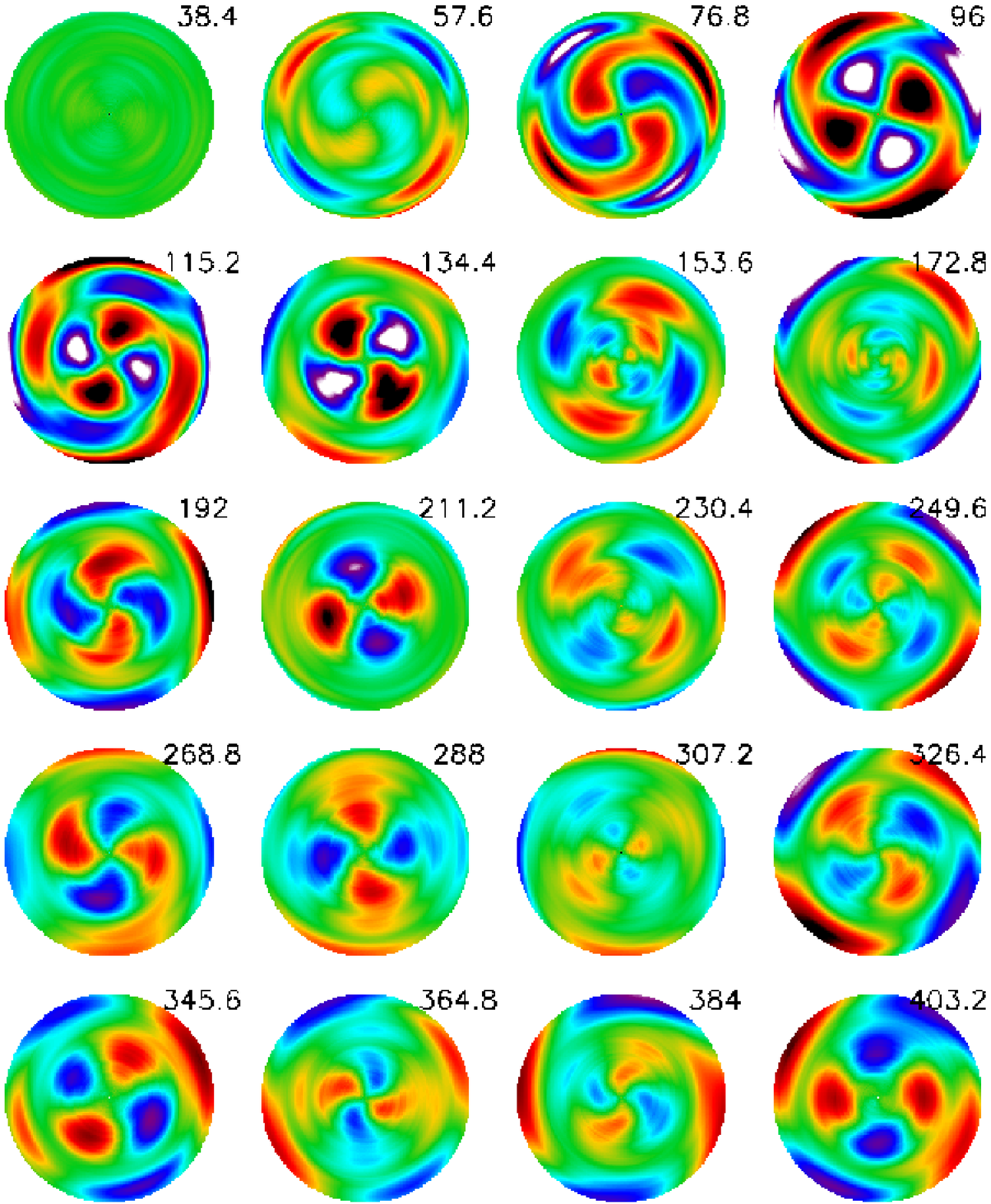}}
\caption[]{{Density contrast associated to $m=2$ mode, for models with $2\times 10^6$ particles in each live component. For $2\times 10^7$ particles the result is qualitatively similar. The color scale ranges over $\pm 0.5$, and indicates the density relative to the mean at each radius. Black and white colors correspond to values outside this range. The radius of each circle is $5 R_d$. The left and right panels belong to models EPL and EM respectively.}}
\label{sectoral}
\end{figure*}

It is clear from Fig. \ref{bar} that the bar grows rapidly in model ERL. On the other hand the growth rate in model EPR is substantially lower than both EM and EPL models. Also there are rapid oscillations in the bra amplitude of the EPR model. However it does not reveal a physical process. In fact for models with rigid components we do not use moving grids. Therefore the center of mass of the system rotates around the grid center and gives rise to these unreal oscillations in the bar growth.

This is a well known fact that disks in live halos form bars readily than in the rigid halos \citep{at2002}. It has been shown in \cite{se16} that the increased growth rate of the bar instability in systems with live halo results from angular momentum exchange between the halo and the disk. We see from the top panel of Fig. \ref{bar} that the exponential growth in both models EPL and EM starts around $\tau \simeq 25$ and reaches its maximum at $\tau \simeq 100$. However the instability growth rate is smaller in the model EM. More specifically, we found that the exponential growth rate, i.e. $e^{\omega \tau}$, in EPL and EM models are $\omega\simeq 0.067$ and $\omega\simeq 0.054$ respectively. 

It is important to mention that there is a significant difference between the bar growth in MOG and MOND. It is shown in \cite{ti} that the growth rate in MOND is substantially higher than the dark matter halo case. It is not the case in MOG, and MOG postpones the occurrence of the bar instability. Albeit although the halo density in \cite{ti} is as ours, the disk density in \cite{ti} is different and given by a Miyamoto-Nagai disk. Therefore, for a more careful comparison between MOG and MOND it is necessary to construct equivalent galactic models with the same surface densities. We leave this  work as a future study.

On the other hand, it is interesting to mention that the maximum value of the bar magnitude in model EM is smaller than in model EPL. This means that MOG leads to weaker bars during the bar instability. This behavior is completely in contrary with MOND which leads to stronger bars during the instability. It is clear in Fig. \ref{bar} that the final bar magnitude in MOG is smaller than EPL. This is also the case in MOND. In other words, although the maximum magnitude of the bar is larger in MOND compared with the dark matter halo model, its final value is smaller \cite{ti}. 

The projected positions of the particles on the $x-y$, $x-z$ and $y-z$ planes for EM and EPL models are shown in Fig. \ref{pos}. The left two columns belong to model EM and two right columns belong to model EPR. In both models, a strong two-arm spiral develops at $\tau\simeq 100$ and rapidly disappear and give place to a bar. The bar in model EPL starts to grow slowly. However, in EM model the bar magnitude decreases and at the same time oscillates.

This oscillation can be related to the existence of some beating between different modes. This oscillatory behavior in model EM can also be seen in the pattern speed evolution. In fact one should note $A_2$ is not only a bar strength, but also a spiral strength. In principle, these waves pattern speeds can be slightly different, and this can give rise to the above mentioned oscillations.

In Fig. \ref{sectoral} we have plotted the relative overdensities associated to $m=2$ mode at different times. The left four columns belong to model EPL, and the right four columns to model EM. In the model EPL, after $\tau \simeq 250$ the bar mode retains its dominance. The twofold symmetry is clear until the end of simulations. On the other hand, in the EM model the evolution is completely different. As it is seen in the right panel of Fig. \ref{sectoral}, the twofold symmetry fades and appears frequently. This is completely in agreement with the period of oscillations observed in Fig. \ref{bar}.
\subsection{Power spectra}
{The clearest way in support of existence of beating between different modes in model EM, is to find the power spectrum. As we already mentioned, we compute the azimuthal Fourier coefficients of the mass distribution at each grid radius and save it at regular time intervals. In order to find the power in terms of frequency and radius for each sectoral harmonic $m$, we compute the Fourier transformation in time of the coefficients.}

\begin{figure*}
\centerline{ \includegraphics[width=16cm]{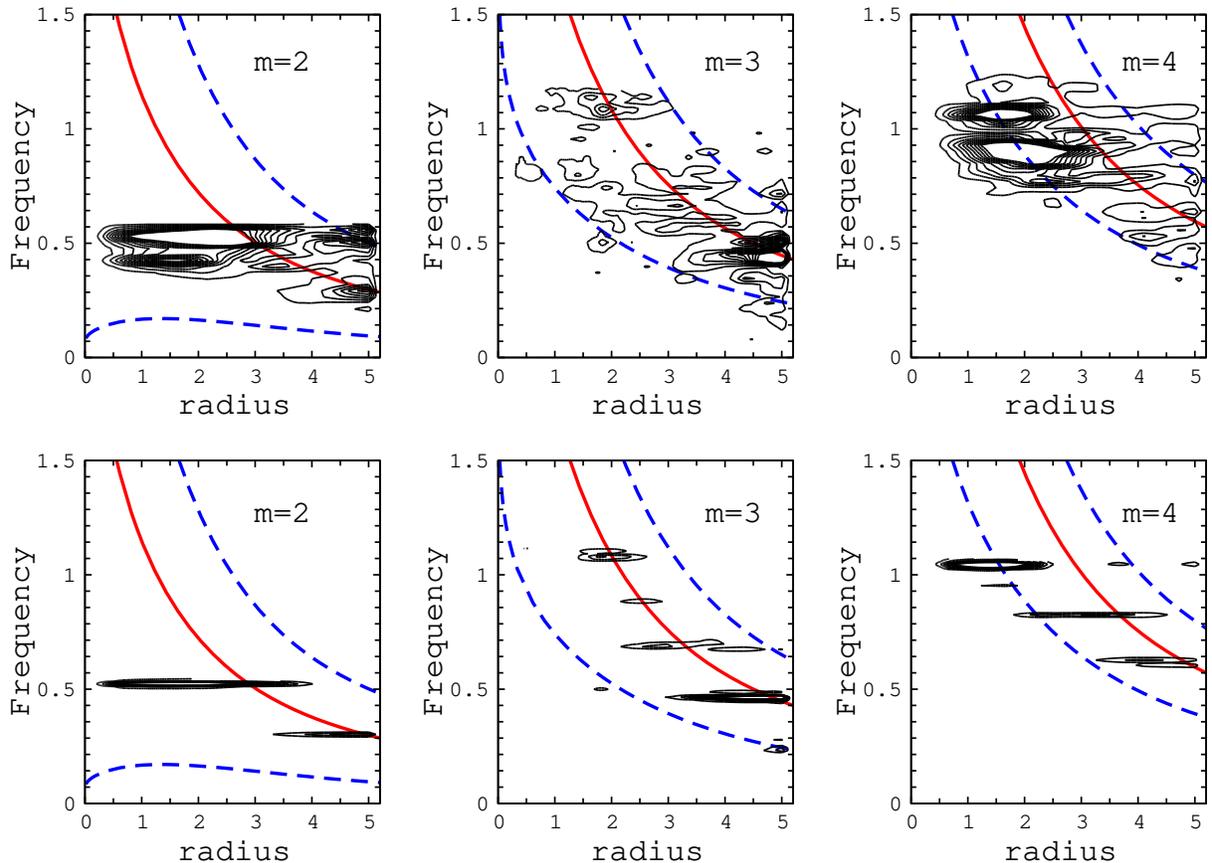}}
\caption[]{{{Contours of power as functions of radius and frequency for different sectoral harmonics $m$ in model EM with $2\times 10^7$ particles. For $2\times 10^6$ particles the results are qualitatively similar. The top row is for the first part of the evolution, i.e. $22 \leq \tau \leq 260$, and the bottom row belongs to the second part of the run. In each panel, the red curve is $m \Omega_c$ and the dashed curves are $m \Omega_c \pm \kappa$. Each horizontal ridge shows a coherent density wave with a well-defined angular frequency, $\Omega_p$, over the specified time interval.}}}
\label{power}
\end{figure*}

{In Fig. \ref{power} we have shown the power spectra for three sectoral harmonics $m=2$, $3$ and $4$ for model EM with $2\times 10^7$ particles. The top row is for the first part of the evolution $22<\tau<260$, and the bottom row is for the second part of the evolution. In each panel, the solid (red) curve marks $m \Omega_c$ and the dashed (blue) curves $m \Omega_c \pm \kappa$. Where $\Omega_c(R)$ is the angular frequency for the circular motion and $\kappa(R)$ is the Linblad epicycle frequency. Since each horizontal ridge indicates a coherent density wave, it is clear from the bottom left panel that there are two $m=2$ waves in the second half of the simulation with pattern speeds $m\Omega_{1}\simeq 0.52 $ and $m\Omega_{2}\simeq 0.3 $. In this case the beat angular frequency is $\Omega_{b}=m(\Omega_{1}-\Omega_{2})= 0.22$. Consequently the beat period is $\tau_b\simeq 28.6$, which is completely in agreement with the period of oscillations in the bar magnitude shown in the bottom panel of Fig. \ref{bar}. Therefore the power spectra confirms the existence of beating between modes.}

{By plotting the magnitude of $m=3$ and $m=4$ modes, it turns out that they also have oscillatory behvior. Silimilar to the two-fold symmetric mode $m=2$, it is clear form the bottom row of Fig. \ref{power} that beating occurs also for $m=3$ and $m=4$ modes. For these modes there are at least four different beating waves with different frequencies.} 

\subsection{Vertical structure}
\begin{figure} 
\centering\includegraphics[width=8.5cm]{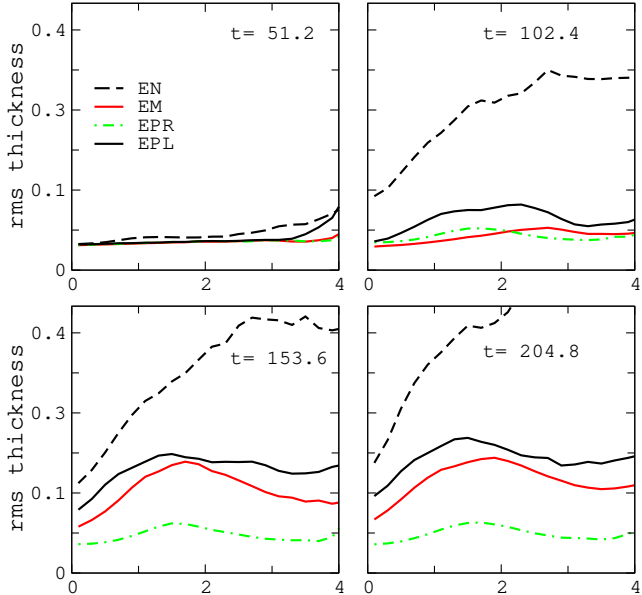}
\caption[]{rms thickness of the disk is plotted with respect to radius for four different times. Although the rms thickness does not grow significantly in EPR, it increases in the models EN, EM and EPL. In this figure we have $2\times 10^6$ particles in each live component. For $2\times 10^7$ particles the result is qualitatively similar.}
\label{thick}
\end{figure}

Particles can resonate both in the plane of disk and perpendicular to it. In this case the vertical oscillations of the particles can be amplified and a peanut shape forms \citep{cs}. One may infer from the top panel of Fig. \ref{bar} that the bar strength drops substantially in both models EPL and EM at $\tau \simeq 150$. On the other hand, at this time the vertical thickness of the disk grows significantly. To see more clearly, we have plotted the root mean square (rms) thickness, i.e. $\sqrt{\langle z^2\rangle}$, in Fig. \ref{thick} for all the models for the same instants presented in Fig. \ref{pos}. Therefore the drop in the bar amplitude coincides with the peanut resonance in both models.  The vertical resonances thicken the disk and consequently stabilize the disk against bar formation. In this case, it is natural to expect that the bar weakens. 
 On the other hand, in the EPR model there is no drop in the evolution of the bar strength, and consequently the thickness does not grow and the peanut shape does not appear. This confirms that the drop in the bar magnitude is due to peanut occurrence. 
 
\begin{figure} 
\centerline{\includegraphics[width=8.5cm]{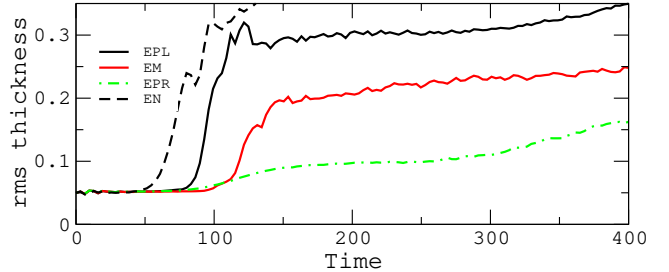}}
\caption[]{Evolution of the rms thickness of the disk with respect to time. The bar instability continuously heats the disk and consequently thickens it. The obvious steps correspond to the occurrence of peanut resonance.}
\label{peanut}
\end{figure}

It is seen in Fig. \ref{thick} that the position of the peanut lobe in both models EM and EPL is in the interval $1.5 \,R_d<R<2 \,R_d$ and does not shift significantly during the disk evolution. Therefore let us measure the rms thickness at $R=1.5\, R_d$ with respect to time. The result has been shown in Fig. \ref{peanut}. The bar instability heats the disk and thickens it continuously. The steps correspond to the peanut occurrence when the particles leave suddenly the plane of the disk. The buckling occurs around $\tau \simeq 80$ in model EPL, and $\tau \simeq 100$ in the EM model. This is consistent with the above mentioned claim that weakening of the bar coincides with the appearance of the peanut.

In both models buckling occurs in a short time interval compared to the simulation time. In model EPL, during the buckling, the rms thickness increases from $0.05\, R_d$ to $0.3\, R_d$. On the other hand in the model EM, the increase rate in slower and the rms thickness increases from $0.05\, R_d$ to $0.2\, R_d$. Therefore this simulation shows that the dark matter stellar model produce stronger peanuts compared with the model in MOG. This is also the case in stellar models in MOND \cite{ti}.

\subsection{Pattern speed $\Omega_p$}
It is instructive to calculate the pattern speed $\Omega_p(t)$. The result has been illustrated in Fig. \ref{ps}. There are $2\times 10^6$ particles in the each live components of the top panel. On the other hand, for comparison, there are $3\times 10^6$ particles in the bottom panel. As reported in \cite{gr2017} the pattern speed in model EM is larger than the EN model. On the other hand, $\Omega_p(t)$ oscillates in the model EM. This is not the case in other models. 

There are some important features in Fig. \ref{ps}. The pattern speed starts with an almost same value for all models. However, although all models begin with same initial conditions, the pattern speed's magnitude and evolution is different when the disks evolve. More specifically $\Omega_p(t)$ oscillates in EM but its mean magnitude is almost constant. However in the model EPL, when the bar is formed at $\tau \simeq 100$, the pattern speed  starts to decrease at $\tau\simeq 125$. In fact the particles of the halo slow down the bar by \textit{dynamical friction}. It is clear that there is no such a decrease in the model EPR where the halo is rigid. So, one may ensure that the substantial decrease in the model EPL is due to the dynamical friction induced by the live halo particles. As expected dynamical friction does not appear in the MOG model, i.e. in model EM. This is the case also in MOND, see \cite{ti} for more details. Absence of dynamical friction in modified gravity can be a crucial point to distinguish between dark matter halo and modified gravity from observational point of view. For example inefficiency of the dynamical friction can substantially reduce the merger frequency of the galaxies. As a consequence, bulge formation and its morphology and kinematics, in principle, can be different from the standard view \citep{com2}.

Let us put emphasize on an important observational fact which is relevant to the pattern speed of disk galaxies. {Observations show that independent of the Hubble type of disk galaxies, bars have been formed and then evolve with time as fast rotators, see \cite{pattern} and references therein.} On the other hand these observations are not in complete agreement with the numerical simulations. For example, a recent $\Lambda$CDM cosmological hydrodynamical simulation by \cite{cosmo} verifies that bars slow down as they grow. Therefore existence of fast bars can be considered as a challenge to $\Lambda$CDM models \citep{deb2000}. 

{In order to compare the pattern speeds in models EM and EPL, let us use the parameter $\mathcal{R}$ defined as}
\begin{equation}
\mathcal{R}=\frac{D_L}{a_B}
\end{equation}
 {where $D_L$ is the corotation radius and $a_B$ is the bar semi-major axis. Bars are fast if $\mathcal{R}\simeq 1$, and is slow if $\mathcal{R}\gg 1$. Finding the corotation radius in the simulation is not hard in the sense that we have the bar pattern speed and the angular frequency $\Omega(R)$ with respect to time. Therefore one may simply estimate the corotation radius. On the other hand in order to estimate the bar semi-major axis at a given time $\tau$, we assume a rectangle with width $\Delta L$ and length $5$ around the line $y = \tan \phi(\tau) x$, where $\phi(\tau)$ is the angular displacement of the bar. Then, we divide this rectangle to small
elements, and choose the bar length to be the length at which the density of the particles is less than $b$ per cent of the central element. We vary $\Delta L$ between $0.2$ and $1$, and choose two values for $b$, i.e. $10$ and $20$. Naturally, these choices lead to an error bar in the final value of $\mathcal{R}$ at each time. }

\begin{figure} 
\centerline{\includegraphics[width=7.5cm]{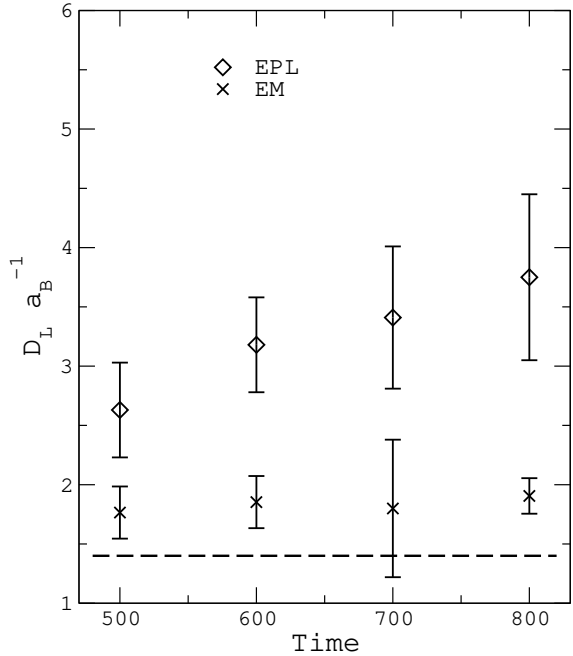}}
\caption[]{{{The ratio $\mathcal{R}=D_L/a_B$ for models EM and EPL with $2\times 10^7$ particles in each live component, calculated at four different times. The error bars are calculated from the averages introduced for measuring the bar length. The dashed line indicates $\mathcal{R}=1.4$. It is clear that bars are faster in modified gravity model than the standard dark matter halo case.} }}
\label{ratio}
\end{figure}

 {The result has been shown in Fig. \ref{ratio} for both models EM and EPL. In this figure we have computed $\mathcal{R}$ for models with $2\times 10^7$ particles in each live component, at four different times. In fact we realize that the corotation radius in model EPL increases with time. More specifically, at $\tau=500$ the corotation radius is $D_L\simeq 3.6$, while at $\tau=800$ it reaches to $D_L\simeq 5.3$. On the other hand $D_L$ stays constant in model EM around $2.7$. Furthermore the bar length for both models is almost constant for $\tau> 500$. Therefore, we expect that $\mathcal{R}$ increases with time in model EPL, and stays constant in model EM. This behavior can be seen in Fig. \ref{ratio}. It is important mentioning that the bar is faster in model EM compared to the dark matter halo model. In fact, in reality most observed bars appear to lie in the range $0.9\lesssim \mathcal{R}\lesssim 1.4$, and consequently observed bars are fast. From this perspective, it is clear from Fig. \ref{ratio} that the model EM is closer to the observed interval, while both models lie outside the interval.}  

However more careful studies are needed to compare MOG with the relevant observations. More specifically, cosmological hydrodynamical simulations would be extremely helpful to perform a statistical study for comparing MOG and observations. To the best of our knowledge, cosmological simulations in the context of alternative theories. which dispense with the need for dark matter particles, have not been developed.  

\section{\small{Discussion and Conclusion}}
\label{disc}
In this paper, we have presented N-body simulations of isolated galaxies in both approaches, of Newtonian gravity with dark matter halo and MOG. The main aim is to compare the bar growth in the presence and absence of dark matter halo. More specifically, in the absence of dark matter halo, we used the modified gravitational force introduced in MOG. We construct three galactic models with almost the same initial conditions, i.e. EM, EPR and EPL. The model EM belongs to MOG and has no spherical component. On the other hand, model EPR is a galactic model with an exponential disk and a rigid Plummer spherical component. The model EPL is more realistic and its halo is a live Plummer sphere. In fact we have a fourth toy model named as EN. This model is an exponential disk in Newtonian gravity without dark matter halo. Comparing this model with EM, we recovered the main results already reported in \cite{gr2017} using low resolution simulations.

Our results show that the bar growth in model EM is slower than in model EPL, and faster than in EPR model. In fact, MOG behaves neither as live dark matter halo nor rigid dark matter halo. However when compared with EN model, it is clear that MOG has stabilizing effects and suppresses the bar instability for a while. Also we found that the final strength of the bar in MOG is smaller than both EPR and EPL models. Moreover the maximum magnitude of the bar in MOG is smaller than the Newtonian models. As we have already mentioned, this is in contrary with MOND where stellar simulations lead to stronger bars. However, although the bar keeps its maximum strength unchanged for a relatively long time in MOND, it keeps its maximum value for a short time in MOG.

We also found that the evolution of the pattern speed in these models is different. More specifically, the pattern speed in model EM, oscillates around an almost constant value and does not change substantially. However the pattern speed in model EPL, decreases significantly by dynamical friction induced by the dark matter particles. In other words, not only the morphology of the bar in our modified gravity theory is different from the standard cases, but also its dynamics is different.{ We explicitly showed that model EM leads to faster bars compared to the dark matter model.}

Furthermore, the rms thickness of the disk in model EM at the end of the simulation is substantially smaller than model EPL. In other words the peanut formation is more effective in model EPL than in model EM. 

More investigations are still required for galactic dynamics in MOG. For example, it would be interesting to study the bar growth in the presence of a massive spherical bulge and galactic gas component. In fact the gas component have crucial effects on the system. For example, bars exert torque on the gas component, which inflows towards the center, and consequently the gas angular momentum is given to the bar. In this case, the bar is weakened, and finally the bar will be faster and shorter \citep{bc}. 

Also, the stabilizing effects of MOG should be compared with other well-known dark matter halo profiles. Furthermore, a careful study is required to compare bar evolution in MOG and MOND. In this case, the initial conditions, i.e. the disk density profile and the velocities, should be the same in both theories. Our results here show that, MOG has some features which are completely different from MOND. We leave these issues as subjects of study for future works.

To summarize the results, let us make a quick comparison between bar evolution in MOG and the dark matter halo models. The main differences are: 1) The bar growth rate is smaller in MOG. 2) The final magnitude of the bar is smaller in MOG. 3) The pattern speed is oscillatory in MOG. Also it does not decrease with time. This is the main difference between MOG and the dark matter models studied in this paper. 4) There is no dynamical friction experienced by the bar in MOG. 5) The maximum value of the bar strength is smaller in MOG. We reiterate that the bar magnitude in MOG stays in its maximum value for a relatively short time interval. In other words, MOG predicts that the strongly barred galaxies are not quiet frequent today. This fact is in agreement with the relevant observations which show that less than 20 \% of barred galaxies are strongly barred \citep{bar1,bar2}.

As our final remark, we stress that in order to make a precise comparison between observation and MOG predictions for the abundance, size and
pattern speeds of bars, as well as for their evolution with redshift, it is necessary to perform high resolution cosmological simulations. For instance for a recent work in the standard cosmological model see \citep{cosmo}. However, there are no similar works in modified theories of gravity designed to dispense with the need for dark matter particles, even in MOND and its relativistic versions. Therefore, a large numerical effort is still required to be able to compare modified gravity predictions for galactic properties with relevant observations.

\acknowledgments
I would like to appreciate Francoise Combes for insightful and constructive comments. {I am particularly grateful to Jerry Sellwood for guidances to run the GALAXY code.} The main part of the calculations have been done on ARGO cluster in International Center for Theoretical Physics (ICTP).  I would like to thank ICTP for a very kind hospitality, during which the main part of this work has been done. This work is supported by Ferdowsi University of Mashhad under Grant No. 2/44972 (18/07/1396).

\bibliographystyle{apj}
\bibliography{paper2}

\begin{thebibliography}{25}
\expandafter\ifx\csname natexlab\endcsname\relax\def\natexlab#1{#1}\fi
\bibitem[{{Aguerri} et al (2015)}]{pattern} Aguerri J. A. L. et al., 2015, A\& A, 576, A102
\bibitem[{{Algorry} et al (2017)}]{cosmo} Algorry D. G. et al., 2017, \mnras, 469, 1054
\bibitem[{{Armengol} \& {Romero} (2017)}]{romero} Armengol, F. G. L. \& Romero, G. E. 2017, arXiv:1611.09918
\bibitem[{{Athanassoula} (2002)}]{at2002} {Athanassoula}, E. 2002, \apj, 569, L83 
\bibitem[{{Athanassoula} (2008)}]{at2008} Athanassoula, E. 2008, \mnras, 390, L69
\bibitem[{{Athanassoula} \& {Selwood} (1986)}]{at1986} {Athanassoula}, E. \&  {Sellwood}, J. A. 1986, \mnras, 221, 213
\bibitem[{{Berrier} \& {Selwood} (2016)}]{bse16} Berrier, J. C. \& Sellwood, J. A. 2016, ApJ, 831, 65
\bibitem[{{Bertone} et al (2010)}]{berton} Bertone, G., Silk, J., Moore, B., et al. 2010, Particle Dark Matter: Observations, Models and Searches (Cambridge: Cambridge Univ. Press)
\bibitem[{{Bournaud} \& {Combes} (2002)}]{bc} Bournaud, F. \& Combes, F. 2002, A\&A, 392, 83
\bibitem[{{Brada} \& {Milgrom} (1999)}]{brada} {Brada}, R. \& {Milgrom}, M. 1999, \apj, 519, 590
\bibitem[{{Brownstein} \& {Moffat} (2006a)}]{br2006} Brownstein, J. R. \& Moffat, J. W. 2006a, \apj, 636, 721
\bibitem[{{Brownstein} \& {Moffat} (2006b)}]{br2006b} Brownstein, J. R. \& Moffat, J. W. 2006b, \mnras, 367, 527
\bibitem[{{Capozziello} \& {De Laurentis} (2011)}]{cap} {Capozziello}, S. \& {De Laurentis}, M. 2011, PhR, 509, 167 
\bibitem[{{Christodoulou} (1991)}]{chris} Christodoulou, D. M. 1991, \apj, 372, 471
\bibitem[{{Combes} (2014)}]{com1} Combes F., 2014, A\&A, 571, A82
\bibitem[{{Combes} (2016)}]{com2} Combes F., 2016, Galactic Bulges, 418, 413
\bibitem[{{Combes} \& {Sanders} (1981)}]{cs} Combes, F. \& Sanders, R. H. 1981, A\& A, 96, 164
\bibitem[{{Debattista} \& {Sellwodd} (2000)}]{deb2000} Debattista, V. P. \& Sellwood, J. A. 2000, \apj, 543, 704
\bibitem[{{de Blok} et al (2008)}]{b2008} de Blok, W. J. G., Walter, F., Brinks, E., Trachternach, C., Oh, S.-H. \& Kennicutt, R. C., Jr. 2008, AJ, 136, 2648
\bibitem[{{Dejonghe} (1987)}]{de} Dejonghe, H. 1987, \mnras, 224, 13
\bibitem[{{D\'{i}az-Garc\'{i}a} et al (2016)}]{bar2} D\'{i}az-Garc\'{i}a, S. and Salo, H. and Laurikainen, E. \& Herrera-Endoqui, M., 2016, A\&A, 587, A160
\bibitem[{{Efstathiou} et al (1982)}]{ef} Efstathiou, G., Lake, G., Negroponte, J. 1982, \mnras, 199, 1069
\bibitem[{{Erwin} (2005)}]{erwin} Erwin P., 2005, \mnras, 364, 283
\bibitem[{{Famaey} \& {McGaugh} (2012)}]{fa} Famaey, B. \& McGaugh, S.S. 2012, LRR, 15, 10
\bibitem[{{Ghafourian} \& {Roshan} (2017)}]{gr2017} Ghafourian, N. \& Roshan, M. 2017, \mnras, 468, no.4, 4450
\bibitem[{{Hohl} (1971)}]{ho} Hohl, F. 1971, \apj, 168, 343.
\bibitem[{{Israel} \& {Moffat} (2016)}]{im} Israel, N. S. \& Moffat, J.W., arXiv:1606.09128 [astro-ph.CO]
\bibitem[{{Jamali} \& {Roshan} (2016)}]{jamali} Jamali, S. \& Roshan, R. 2016, EPJC, 76, 490 
\bibitem[{{Jamali} et al (2017)}]{jamali2} Jamali, S., Roshan, R. \& Amendola, L. 2017, arXiv:1707.02841
\bibitem[{{Milgrom} (1983)}]{milgrom} Milgrom, M. 1983, \apj, 270, 384
\bibitem[{{Milgrom} (1989)}]{mil89} Milgrom, M. 1989, \apj, 338, 121
\bibitem[{{Miller} et al (1970)}]{miller} Miller, R. H., Prendergast, K. H., Quirk, W. J. 1970, \apj, 161, 903
\bibitem[Moffat (2006)]{m2006} Moffat, J. W. 2006, JCAP, 0603, 004
\bibitem[Moffat (2015a)]{m2015} Moffat, J. W. 2015a, EPJC, 75, 130 
\bibitem[Moffat (2015b)]{gravitoelectric} Moffat, J. W. 2015b, EPJC, 75, 175
\bibitem[Moffat \& Rahvar (2013)]{m2013} Moffat, J. W. \& Rahvar, S. 2013, \mnras, 436, 1439
\bibitem[Moffat \& Toth (2009)]{m2009} Moffat, J. W. \& Toth, V. T. 2009, CQG, 26, 085002
\bibitem[Monaghan (1992)]{mon1992}  Monaghan, J. J. 1992, ARAA, 30, 543
\bibitem[Ostriker \& Peebles (1973)]{op} Ostriker, J. P. \& Peebles, P. J. E. 1973, \apj, 186, 467
\bibitem[Roshan \& Abbassi (2014)]{ro2014} Roshan, M. \& Abbassi, S. 2014, PhRvD, 90, 044010 
\bibitem[Roshan \& Abbassi (2015)]{ro2015} Roshan, M. \& Abbassi, S. 2015, \apj, 802, 9 
\bibitem[Roshan et al (2016)]{rokho} Roshan, M. \& Abbassi, S. \& G. Khosroshahi H. 2016, \apj, 832,
no.2, 201
\bibitem[Roshan (2015)]{roepjc} Roshan, M. 2015, EPJC, 75, 405
\bibitem[Saha \& Naab (2013)]{saha} Saha, K., Naab, T. 2013, \mnras, 434, 1287
\bibitem[{{Sellwood} (1981)}]{se81} Sellwood, J. A. 1981, \aap, 99, 362
\bibitem[{{Sellwood} (2003)}]{se3} Sellwood, J. A. 2003, \apj, 587, 638
\bibitem[{{Sellwood} (2011)}]{se11} Sellwood, J. A. 2011, \mnras, 410, 1637
\bibitem[{{Sellwood} (2014a)}]{se2014} Sellwood, J. A. 2014a, arXiv:1406.660
\bibitem[{{Sellwood} (2014b)}]{se14} Sellwood, J. A. 2014b, RvMP, 86, 1
\bibitem[{{Sellwood} (2016)}]{se16} Sellwood, J. A. 2016, \apj, 819, 92
\bibitem[{{Sellwood} \& {McGaugh} (2005)}]{sm2005} Sellwood, J. A. \& McGaugh, S. S. 2005, \apj, 634, 70
\bibitem[{{Shojai} et al (2017)}]{shojai} Shojai, F., Cheraghchi, S. \& Bouzari Nezhad, H. 2017, PLB, 770, 43
\bibitem[{{Tiret} \& {Combes} (2007)}]{ti} Tiret, O., \& Combes, F. 2007, \aap, 464, 517
\bibitem[{{Toomre} (1964)}]{toom64} Toomre, A. 1964, \apj, 139, 1217
\bibitem[{{Whyte} et al (2002)}]{bar1} Whyte, L. F., Abraham, R. G., Merrifield, M. R., et al. 2002, \mnras, 336, 1281
\bibitem[{{Young} (1980)}]{y1980} Young, P. 1980, \apj, 242, 1232
\end{thebibliography}
\label{lastpage}
\end{document}